\documentstyle[preprint,pra,oldlfont,aps]{revtex}

\newcommand{\pmf}{\phi_{mf}(s)}
\newcommand{\psmf}{\phi_{mf}^2(s)}
\newcommand{\N}{\overline{N}}
\newcommand{\x}{{\bf x}}
\newcommand{\y}{{\bf y}}
\newcommand{\xa}{{\bf x}^{\alpha}}
\newcommand{\ya}{{\bf y}^{\alpha}}
\newcommand{\hx}{\widehat{{\bf x}}}
\newcommand{\pso}{\phi_1^2(\hat{{\bf x}})}
\newcommand{\pst}{\phi_2^2(\hat{{\bf x}})}
\newcommand{\po}{\phi_1(\hat{{\bf x}})}
\newcommand{\ppt}{\phi_2(\hat{{\bf x}})}
\newcommand{\ffo}{\psi_1(\hat{{\bf x}})}
\newcommand{\fft}{\psi_2(\hat{{\bf x}})}
\newcommand{\fso}{\psi_1^2(\hat{{\bf x}})}
\newcommand{\fst}{\psi_2^2(\hat{{\bf x}})}
\newcommand{\tha}{\theta_{\alpha}(\xa)}
\renewcommand{\ffo}{\psi_1(\hat{{\bf x}})}
\renewcommand{\fft}{\psi_2(\hat{{\bf x}})}
\newcommand{\ma}{m_{\alpha}(\xa)}
\newcommand{\msa}{m_{\alpha}^2(\xa)}
\newcommand{\q}{{\bf q}}
\newcommand{\qa}{{\bf q}^{\alpha}}
\newcommand{\mqa}{m_{\q^{\alpha}}}
\newcommand{\msqa}{m_{\q^{\alpha}}^2}
\newcommand{\nqa}{{\bf n}_{\q^{\alpha}}}
\newcommand{\mnqa}{m_{-\q^{\alpha}}}
\newcommand{\nnqa}{{\bf n}_{-\q^{\alpha}}}
\newcommand{\n}{{\bf n}}
\newcommand{\gq}{g_{\qa}}

\begin{document}

\draft
\title{Statistical mechanics of a cross-linked polymer blend}
\author{C. D. Sfatos, A. M. Gutin and E. I. Shakhnovich}
\address{Harvard University, Department of Chemistry\\
12 Oxford Street, Cambridge Massachusetts 02138}
\date{\today}
\maketitle
\begin{abstract}
We study a blend of two kinds of homopolymers with tendency
for segragation. Cross-links between the chains of different
kinds do not allow macrophase separation. Instead microphase
structure appears.
Starting from a microscopic model we derive the effective Hamiltonian
and calculate the form of the inverse scattering function
and the domain size. The latter is found to be of the order of the
mesh size. 
We show agreement of the results obtained by this
microscopic statistical mechanical theory with experimental 
data and the existing scaling arguments.
\end{abstract}

Polymer gels exhibit a wide spectrum of interesting properties 
resulting from their mechanical stability, their elastic properties
 and their ability to swell upon absorbtion of solvent, sometimes
giving rise to a multitude of swollen phases \cite{li-tanaka}. The theoretical
approach to such systems with cross-link constraints has
also been related to the study of the determination of protein
structures given a set of constraints produced by protein NMR
methods \cite{gutshak}.

The statistical mechanics of systems with random cross-links 
have attracted significant attention and have been the subject
of extensive debate in the interpetation of the results obtained by
the different methods 
\cite{deam-edwards,gold1,gold2,zippel,pan-rab}. 
In order to take into account
the randomness in the distribution of the cross-links the replica
method was used and the interpretation of the multi-replica
order parameter was lacking. Recently, Panyukov and Rabin \cite{pan-rab}
derived a physical effective Hamiltonian that allows
the calculation of observable parameters such as density correlation
functions due to randomness in the quenched
cross-links and due to thermal fluctuations. Besides this seminal
contribution, their theory is also consistent with the behavior
of thermal and quenched fluctuations due to deformation (butterfly patterns)
\cite{bastide,rouf}.

In the present work we apply the Panyukov-Rabin method to derive
the effective Hamiltonian of a system of a blend of 
two very long homopolymeric chains
of different kinds that are cross-linked 
\cite{degennes,stepanow,daoud}. We consider the case where
there is an effective repulsion between the two kinds of monomers
that leads the uncross-linked  blend to segragation at low temperatures.
We assume that the cross-linking takes place at high temperatures
where the blend is in the mixed state. After the system is cross-linked
reduction of temperature will not lead to macroscopic segregation because
the cross-links hold the pieces of different kinds together. Instead
a microphase separation will appear.

The study of this transition so far has been based
on a phenomenological Hamiltonian introduced by de Gennes \cite{degennes} where
the effect of cross-links has been taken into account as a Hook-law
type of elastic forces that keep the pieces of different kinds
together. A complete derivation of the effective Hamiltonian
starting from a microscopic model, to the best of our knowledge
was missing. The result obtained here provides a deep understanding
of the physics of the system and its quantitative predictions are
in fine agreement with experimental data as discussed below \cite{briber}.

In a microscopic description we consider a Gaussian chain with
Kuhn segment $a$ described in the Edwards formulation \cite{deam-edwards} by
\begin{equation}
{\cal H}_0=\frac{3}{2a^2}\int_0^N di \left(\frac{d\x}{di}\right)^2
+\frac{w^{(0)}}{2}\int_0^Ndi\int_0^Ndj \delta [\x (i)-\x (j)],
\end{equation}
where $w^{(0)}$ the excluded volume.
In our problem we have two kinds of chains, e.g., one white
chain and one black, that have a tendency for segregation.
We denote the position of the monomers of one kind by $\x (i)$
and the other by $\y (i)$. The interaction between the two kinds
is given by

\begin{equation}
{\cal H}_1=-\chi\int di\int dj \delta [\x (i)-\y (j)].
\end{equation}

 The cross-link constraints
are introduced by a set of delta-functions and the partition
function can be written as
\begin{equation}
Z=\int{\cal D}\x (i)
\exp\Bigl[-{\cal H}_0(\x (i))-{\cal H}_0(\y (i))
-{\cal H}_1(\x (i), \y (i))
\Bigl]
\prod_{p=1}^{N_c}
\delta[[\x (i_p)-\y (j_p)],
\end{equation}
where $p$ is a label for a particular cross-link and $N_c$ the number
of cross-links in a particular realization of disorder.

Then we use the replica trick \cite{binder} to perform the average
over all realisations of cross-links. However, taking this 
disorder average it is important to impose an additional constraint:
that the cross-links are compatible with an existing real conformation
\cite{gutshak,deam-edwards,pan-rab}.
To ensure this
we fix  the cross-links to be compatible with a conformation
labeled as an additional ``zeroth'' replica. 
Also with this constraint we ensure that the number of 
ways to chose $N_c$ cross-links should not
exceed the total number of available conformations 
 and the conformations over which the 
partition function average is taken are physical. Then the $n$-th power
of the partition function averaged over disorder reads:
\begin{eqnarray}
\langle Z^n\rangle=&&
\int d{\bf S} \prod_{\alpha=0}^n
\biggl\{
\int {\cal D}\xa (i)\int {\cal D}\ya (i)
\exp\Bigl[ -{\cal H}_0(\xa (i))-{\cal H}_0(\ya (i))
-{\cal H}_1(\xa (i), \ya (i))\Bigl]
\nonumber\\
&&\times\prod_{p=1}^{N_c}
\delta[\xa (i_p)-\ya (j_p)]
\biggl\},
\end{eqnarray}
where the average over all cross-link configurations is given by
\begin{equation}
\int d{\bf S}\equiv \frac{1}{N_c!}
\prod_{p=1}^{N_c}\int_0^N di_p\int_0^N dj_p.
\end{equation}

The derivation of the effective Hamiltonian from this
model can be done by using the grand canonical ensemble
with a fixed number of $N$ monomers for each chain. The
first order parameter that appears in this process is
\begin{equation}
\rho_1(\hx)=\rho_1(\x^0, \x^1,\ldots ,
\x^{\alpha},\ldots ,\x^n)=
\int_0^N\prod_{\alpha =0}^n
\delta (\xa (i)-\xa),
\end{equation}
where $\hx$ is a multi-replica $3(n+1)$-dimensional probe vector,
$\xa (i)$ is the position of monomer $i$ of replica $\alpha$,
and $\xa$ is a probe position in the 3-dimensional space of replica
$\alpha$. The corresponding parameter $\rho_2 (\hx)$ is defined
for the monomers of the other kind. This parameter can be interpreted
as the average correlator of conformations of the $n$ replicas
with the ``zeroth'' replica which corresponds to an existing conformation
at the moment when cross-linking takes place. This conformation
is also taken to be maximally compact with constant real density.

In order to derive the effective Hamiltonian for our system
we need to calculate the entropy of conversion from the microscopic
to the macroscopic parameter with introduction of the fields
$\phi_1^2(\hx)/2=\rho_1(\hx)$ and $\phi_2^2(\hx)/2=\rho_2(\hx)$
\cite{pan-rab,gutshak}.
The part of the effective Hamiltonian that does not contain the repuslive
interaction between monomers of different kinds ${\cal H}_1$, then reads
\begin{eqnarray}
{\cal H}_0[\phi_1, \phi_2]=&&
\int d\hx\biggl\{
\frac{1}{2}(\mu_1\pso+\mu_2\pst)
+\frac{a^2}{2}\Bigl[
(\hat{\nabla}\po)^2+(\hat{\nabla}\ppt)^2
\Bigl]\nonumber\\
&&-\frac{z_c}{4}
\pso\pst
+\frac{w^{(0)}}{2}\sum_{\alpha =0}^n
\int d\xa
\Bigl[\int
\Bigl(\prod_{\beta\neq\alpha}
d\x^{\beta}\Bigl)
(\pso+\pst)
\Bigl]^2
\biggl\}.
\end{eqnarray}
The chemical potentials $\mu_1, \mu_2$ are introduced in the grand canonical
ensemble formalism to fix the number of monomers of each chain to $N$. They 
are equal with each other and are given by
\begin{equation}
\mu=\frac{1}{\N}-w^{(0)}\rho,
\end{equation}
where $\N$ is the average number of monomers from the one 
cross-link to the next and $\rho$ the mean field density.

The number of cross-links is proportional to $z_c$ which
is given by
\begin{equation}
z_c=\frac{1}{\rho \N}.
\end{equation} 

The excluded volume effect is important for the ``zeroth'' replica
since it fixes the maximal number of cross-links at the cross-link
saturation threshold $w^{(0)}\rho\N=1$.

This Hamiltonian and the corresponding multi-replica order parameter
do not give appreciable physical insight and some physical order parameters
should be sought. Following the methodology of Panyukov and Rabin
we perform a procedure for the evaluation of the entropy for the conversion
into an order parameter related to differnce of the local densities
of the two kinds.

Originaly we observe that the mean field values for the densities
of the two kinds of monomers should be equal because of symmetry.
It can be shown \cite{pan-rab} that this mean field solution is given by a one
variable mean field equation
\begin{equation}
\Biggl(
\frac{1}{\N}-\frac{2a^2s}{3}\frac{\partial^2}{\partial s^2}
-z_c\psmf
\Biggl)\pmf=0,
\end{equation}
where 
\begin{equation}
s=\frac{1}{2}
\biggl[
\Bigl(\sum_{\alpha}\xa\Bigl)^2
-\frac{1}{n+1}\sum_{\alpha}(\xa )^2
\biggl].
\end{equation}
The single variable $s$ of this differential equation
corresponds to the average deviation of the $n$ replica
conformations from the ``zeroth'' replica, i.e., the initial conformation
at the moment of cross-linking. The solution of the mean field equation
can be done numerically.

However, the local fluctuations should account for the differences
that give rise to a microphase separation pattern. The microphase
separation parameter should be calculated as a difference between
the single replica fluctuation differences. The latter can be calculated
from
\begin{equation}
\rho_1^{\alpha}(\xa)=\int
\Bigl(\prod_{\beta\neq\alpha}
d\x^{\beta}\Bigl)
\pso\ \ \ \ ;\ \ \ \ 
\rho_2^{\alpha}(\xa)
=\int\Bigl(\prod_{\beta\neq\alpha}
d\x^{\beta}\Bigl)\pst .
\end{equation}
If we set $\po =\po +\ffo$ and $\ppt =\ppt +\fft$, where $\ffo,\fft$ 
the fluctuations in the multi-replica field, the lowest order 
estimate with respect to fluctuations
for the difference between densities of the two kinds is given
by
\begin{equation}
2\int\Bigl(\prod_{\beta\neq\alpha}
d\x^{\beta} \Bigl)
(\ffo-\fft)\pmf .
\end{equation}
We need to evaluate the entropy for the conversion to the physical
single-replica order parameters that correspond to the microphase
separation in real space. This can be done with the introduction
of an auxiliary field $\theta$ and the $n$-th power of the partition
function which reads
\begin{eqnarray}
\langle Z^n\rangle=&&
\int {\cal D}\ma{\cal D}\tha
\exp\Bigl[
-\chi\sum_{\alpha}\int d\xa\msa+
\sum_{\alpha}\int d\xa\ma\tha
\Bigl]
\nonumber\\
&&\times\biggl\langle
\exp \Bigl\{
-2i\int d\hx \pmf
[\ffo-\fft ]
\sum_{\alpha}\tha
\Bigl\}
\biggl\rangle_{\Delta{\cal H}},
\end{eqnarray}
where $\ma$ is the macroscopic microphase separation  
order parameter  as a result of the difference
between fluctuations of the densities of the two kinds
of monomers and $\Delta{\cal H}$ is the part of the
Hamiltonian that depends on these density fluctuations, given by

\begin{eqnarray}
\Delta {\cal H}[\ffo ,\fft]=&&
\frac{1}{2}\int d\hx
\Bigl[ \frac{1}{\N}
(\fso +\fst )\Bigl]
-\frac{a^2}{3}[\nabla^2\ffo +\nabla^2\fft ]
\nonumber\\
&&-\frac{z_c}{2}\Bigl[
\fso +\fst \Bigl]\psmf
-2z_c\ffo
\fft\psmf .
\end{eqnarray} 

The evaluation of this conversion entropy results
into a more physical expression for the effective
Hamiltonian which reads

\begin{equation}
\langle Z^n \rangle =
\Biggl\langle
\int {\cal D}\mqa
\exp \biggl\{ -\frac{1}{2}\sum_{\alpha=1}^n
\Bigl[ \int d\qa
\frac{(\mqa-\nqa)
(\mnqa-\nnqa)}{\gq}
-\chi \int d\q \msqa
\Bigl]\biggl\}\Biggl\rangle_{\n}.
\label{final}
\end{equation}
where $\mqa$ is the Fourier transform of the microphase 
separation order parameter.
At this stage we see that there is no coupling between replicas
so that the replica index becomes irrelevant and can be droped.
The newly introduced field $\n$
corresponds to the quenched local differences of the
densities of the two kinds of monomers associated with the initial
conformation. The deviation from this random quenched
microphase structure raises the energy due to the network
elastic forces. The form of the correlation function
will depend on the thermal correlation function $g_{\q}$ 
and will be discussed below. The field $\n$ defines
the profile for a reference state with fixed density.
This state is very different
from an ideal reference state without excluded volume.
which is  collapsed to the dimensions of the mesh
size and cannot serve as reference to the excluded
volume problem problem of an incompressible melt.
Although an elegant way has been recently proposed for the ideal
chain with cross-links \cite{vilgis} this cannot serve as a basis for the
solution of the incompressible, real melt.

Thermal fluctuations correspond to probing a particular
part of space in a gel with a given set of cross-links
and corresponding to a time average. The average over different
disorder realizations, i.e., different sets of cross-links,
should be done with the aid of the probability distribution
of the field $\n$ which can be calculated according to the 
principles  laid by Panyukov and Rabin and will be presented elsewhere
since, as it will be discussed below, it is not important for the study 
of the microphase separation transition principles.

The most important quantity is $g_{\q}$ which can be calculated
by a long procedure that involves numerical solutions of differential
equations and integrals and will be presented in detail elsewhere.
The result obtained for the large scale limit $\q\rightarrow 0$ is
$g(\q\rightarrow 0)=0.60a^2\N^2(\rho)^{1/2}\q^2$. 
This expession is different from the corresponding correlation
function for the density fluctuations found by Panyukov and Rabin
in that the leading term is proportional to $\q^2$ and not of order unity.

For short scale behavior, i.e.,
at scales shorter than the mesh size where the system does not feel
the cross-links, the density fluctuation correlation function is
given by $G_0(\q)=12/(a^2\q^2)$ for each of the two chains
of our system \cite{book}. By setting $\rho_1=(1/2)(\rho +m)$ and 
$\rho_2=(1/2)(\rho -m)$ we can find easily  the contribution 
of this term to the free energy, so that the effective Hamiltonian
to second order becomes
\begin{equation}
{\cal H}=\frac{1}{2}\int d\q
\Biggl[
\frac{1.66}{a^2\N^2(\rho)^{1/2}\q^2}+
\frac{a^2}{24}\q^2+\frac{1}{2}(\chi-\chi_0)
\Biggl]m_{\q}^2,
\label{correl}
\end{equation}
where $(\chi-\chi_0)$ is the shift from the temperature of segregation
for the uncross-linked blend. The prefactor 1.66 differs from the estimate
36 of de Gennes. The fourth order vertex is much more difficult
to calculate and is an open question.

The inverse correlation function of Eq.\ (17) gives rise to
a Hamiltonian of the Brazovskii \cite{braz}
type $A(q-q_0)^2+\tau$ with a minimum at
\begin{equation}
aq_0=\frac{2.5}{\N^{1/2}\rho^{1/8}},
\label{min}
\end{equation}
and the density can be set equal to unity for simplicity.
This coincides with the result obtained by de Gennes
appart from a numerical prefactor. The correlation function
diverges for $q=q_0$ at the transition point $\tau =0$, whereas the
quenched fluctuations do not depend on temperature, i.e., the proximity
to the transition point, so they will not be important for
the observations near the transition point.

The prefactor 2.5
should be considered more accurate, because it comes from 
a microscopic theory and is in agreement with the experimental
value of 2.3 which is found from the microphase separation
wave length mesurements in a blend of deuterated polystyrene
with poly(vinyl methyl ether) crosslinked with radiation \cite{briber}.

With a statistical mechanical treatment for gels of the sort
described here and the encouraging results obtained in this work
the way is now open to tackle more chalenging problems
in the filed of thermodynamics
of heteropolymeric gels such as the case which includes
electrostatic forces and which shows complex behavior and multiple phases
\cite{annaka} that have remained an open problem in the field for along time.


\begin{thebibliography}{10}

\bibitem{li-tanaka}
Y. Li and T.Tanaka, Annu.\ Rv.\ Mater.\ Sci.\
{\bf 22}, 243 (1993).

\bibitem{gutshak}
A. M. Gutin and E. I. Shakhnovich, J. Chem.\ Phys.\ {\bf 100}, 5290 (1994).

\bibitem{deam-edwards}
R. T. Deam and S. F. Edwards, Philos.\ Trans.\ R. Soc.\ London Ser.\ A
{\bf 280}, 317 (1976).
 
\bibitem{gold1}
P. M. Goldbart and N. Goldenfeld, Phys. Rev.\ Lett.\
{\bf 58}, 2676 (1989).

\bibitem{gold2}
P. M. Goldbart and N. Goldenfeld, Phys. Rev.\ A {\bf 39}, 1402 (1989).

\bibitem{zippel}
A. Zippelius, P. M. Goldbart, and N. Goldenfeld,
Europhys.\ Lett.\ {\bf 23}, 451 (1993).

\bibitem{pan-rab}
S. Panyukov and Y. Rabin, preprint (1995).

\bibitem{bastide}
J. Bastide, L. Leibler, and J. Prost, Macromolecules {\bf 23}, 1821 (1990).

\bibitem{rouf}
C. Rouf, J. Bastide, J. M. Pujol, F. Schosseler, and J. P. Munch,
Phys.\ Rev.\ Lett.\ {\bf 73} 830 (1994).

\bibitem{degennes}
P. G. de Gennes, J. Phys.\ Lett.\ (Paris) {\bf 40}, 69 (1979).

\bibitem{stepanow}
S. Stepanow, M. Schulz, and K. Binder, J. Phys.\ II (Paris) 
{\bf 4}, 819, (1994).

\bibitem{daoud}
A. Betachy, A. Derouiche, M. Benhamou, and M. Daoud,
J. Phys. I (Paris) {\bf 1}, 153 (1991).

\bibitem{binder}
K. Binder and A. P. Young, Rev.\ Mod.\ Phys.\ {\bf 58}, 801 (1986).

\bibitem{vilgis}
M. P. Soft and T. A. Vilgis, J. Phys.\ A {\bf 28}, 6655 (1995).

\bibitem{book}
P. G. de Gennes, {\em Scaling Concepts in Polymer Physics} (Cornell University
  Press, Ithaca {NY}, 1979).

\bibitem{braz}
S. A. Brazovskii, Zh.\ Eksp.\ Teor.\ Fiz.\ {\bf 68}, 175 (1975)
[Sov.\ Phys.\ JETP {\bf 41}, 85 (1975)].

\bibitem{briber}
R. M. Briber and B. J. Bauer, Macromolecules {\bf 21}, 3296 (1988).

\bibitem{annaka}
M. Annaka and T. Tanaka, Nature {\bf 355}, 430 (1992).

\end{thebibliography}
\end{document}